\documentclass[12pt]{article}
\usepackage{graphicx}
\usepackage{wrapfig}
\usepackage{setspace}
\usepackage{subfig}
\newcommand\pubnumber{SNSN-323-63}
\newcommand\pubdate{\today}

\def\institute{Department of Nuclear Physics and Biophysics\\
Comenius University, Bratislava, Slovakia}
\def\support{\footnote{Work supported by the Ministry of Education, Science, Research and Sport of the Slovak
Republic.}}

\def\Title#1{\begin{center} {\Large #1 } \end{center}}
\def\Author#1{\begin{center}{ \sc #1} \end{center}}
\def\Address#1{\begin{center}{ \it #1} \end{center}}

\newcommand\pubblock{\rightline{\begin{tabular}{l} \pubnumber\\
         \pubdate  \end{tabular}}}
\newenvironment{Abstract}{\begin{quotation}  }{\end{quotation}}
\newenvironment{Presented}{\begin{quotation} \begin{center} 
             PRESENTED AT\end{center}\bigskip 
      \begin{center}\begin{large}}{\end{large}\end{center} \end{quotation}}

\def\beq{\begin{equation}}
\def\eeq#1{\label{#1}\end{equation}}
\def\eeqn{\end{equation}}

\def\beqa{\begin{eqnarray}}
\def\eeqa#1{\label{#1}\end{eqnarray}}
\def\eeqan{\end{eqnarray}}

\let\bar=\overbar

\def\Dslash{\not{\hbox{\kern-4pt $D$}}}
\def\dslash{\not{\hbox{\kern-2pt $\del$}}}

\def\msb{{\bar{\ssstyle M \kern -1pt S}}}

\begin{document}
\begin{titlepage}
\pubblock

\vfill
\Title{Top Quark Properties at the Tevatron }
\vfill
\Author{ Stanislav Tokar\support}
\Address{\institute}
\vfill
\begin{Abstract}
I report on the status of the studies of top-quark properties carried out by the Tevatron experiments CDF and D0.
\end{Abstract}
\vfill
\begin{Presented}
$9^{th}$ International Workshop on Top Quark Physics\\
Olomouc, Czech Republic,  September 19--23, 2016
\end{Presented}
\vfill
\end{titlepage}
\def\thefootnote{\fnsymbol{footnote}}
\setcounter{footnote}{0}

\section{Introduction}

The Tevatron accelerator finished its operation in 2011. Working with colliding beams of protons and anti-protons ($p\bar{p}$) at a center-of-mass energy of $\sqrt{s}$ = 1.96~TeV the Tevatron experiments, CDF \cite{CDF_det} and D0 \cite{D0_det}, collected data samples with an integrated luminosity of 10 fb$^{-1}$ per experiment.
The contribution of the Tevatron experiments to the particle physics is manifold. Above all it was the discovery of the top quark and  determination of its basic properties. In addition there were accomplished such discoveries as the discovery of $B^0_\mathrm{S}$-oscillation, other significant measurements of heavy quark production, jet physics and electroweak physics. 
The contribution  of the Tevatron experiments is also enormous to the experimental techniques presently used in particle physics. 
Though the Tevatron accelerator was stopped, results of the data analysis are coming and they are still competitive.




\vspace*{-6pt}
\section{Results}
The Tevatron experiments studied many different aspects of top-quark physics. Among the studied topics were (and still are) not only the total and differential cross sections, the top-quark mass, but also the production asymmetries, spin correlations, top-quark polarization, top-quark width, top-quark charge, anomalous couplings, helicity of $W$ bosons from top-quark decay and some other topics. Only some of these topics are mentioned in this contribution.
\vspace*{-12pt}
\subsection{The top-quark pair production cross section}
The theoretical $t\bar{t}$ production cross section,  the total and differential one, is now known  
\begin{wrapfigure}{r}{0.40\textwidth}
\vspace{-3mm}
\centering
\begin{tabular}{c}
\includegraphics[width=0.39\textwidth]{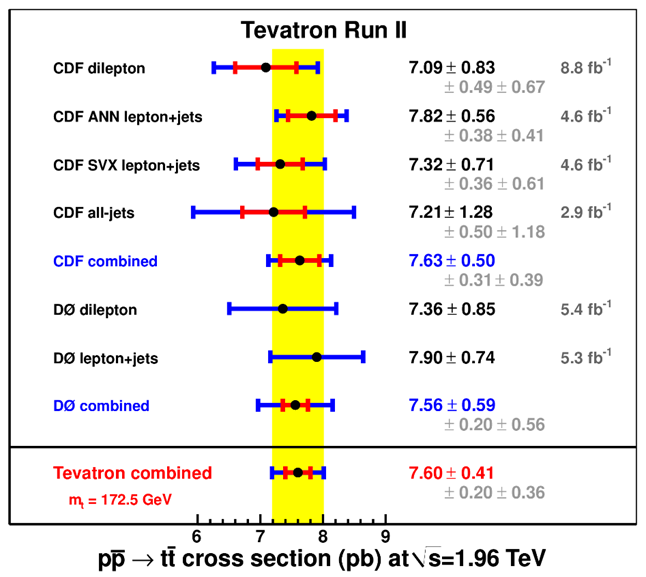}
\end{tabular}
\vspace{-5mm}
\caption{{\small Tevatron $t\bar{t}$ production cross section summary.}}
\label{fig:Xs_summary}
\vspace{-7mm}
\end{wrapfigure}
at the next-to-next-to-leading order 
(NNLO) complemented with the next-to-next-to-leading logarithmic (NNLL) soft gluon resumation \cite{czakon_difXs}. 
An inclusive $t\bar{t}$  cross section measurement carried out by D0 is shown as an example of a Tevatron cross section study. The measurement combines lepton+jets ($\ell$+jets) and dilepton ($\ell\ell$) channels \cite{D0_Xs} 
using a data sample of 9.7~fb$^{-1}$.
The cross section was extracted using a combined likelihood which is a product of binned likelihoods for the individual channels. The measured combined cross section is:
\begin{center}
\vspace{-2mm}
$\sigma_{t\bar{t}} =$ 7.26 $\pm $ 0.13 (stat) $^{+0.57}_{-0.50}$ (syst) pb
\vspace{-2mm}
\end{center}
The systematic uncertainty dominates and the  measurement relative precision is 7.6\%.
The results of the CDF and D0 cross section measurements were combined giving:
\begin{center}
\vspace{-2mm}
$\sigma_{t\bar{t}} =$ 7.60 $\pm $ 0.41 pb
\vspace{-2mm}
\end{center}
with the relative precision of 5.4\%. The input measurements including the combined cross section are shown in Fig.\ref{fig:Xs_summary}. 
It should be noted that the latest D0 result is not included in the combined result.  The final CDF result is under preparation.   
\vspace*{-12pt}
\subsection{Asymmetries in the top-quark pair production}
Under the standard model (SM), the asymmetries in $t\bar{t}$ production arise only in higher orders of $q\bar{q}$ annihilation, and quark-gluon flavor excitation while gluon-gluon fusion is symmetric in all orders \cite{prd59_khun}. 
The $t\bar{t}$ production asymmetries are investigated using angle or rapidity 
distributions of the reconstructed $t\bar{t}$ pair or the leptons from the top-quark decays.

\noindent {\bf Forward-backward asymmetry -- direct approach.} The forward-backward (FB) asymmetry is measured using the difference of
the $t$ and $\bar{t}$ rapidities, $\Delta y = y_{t}-y_{\bar{t}}$, and is expressed through the number of events ($N$) with positive and negative $\Delta  y$:
\vspace*{-8pt}
\begin{eqnarray}
A_\mathrm{FB}^{t\bar{t}} = \frac{N(\Delta y>0)-N(\Delta y<0)}{N(\Delta y>0)+N(\Delta y<0)}. 
\label{eq:AFB}
\end{eqnarray}
\vspace*{-15pt}

\noindent CDF measured the FB asymmetry in the $\ell+$jets and $\ell\ell$ channels using the  data sets of 9.4~fb$^{-1}$ \cite{prd87_cdf_afb}  and 9.1~fb$^{-1}$ \cite{prd93_cdf_afb-ll}, respectively. 
The extracted asymmetries unfolded to parton level 
were combined (see Ref. \cite{prd93_cdf_afb-ll}), resulting in 
\vspace*{-6pt}
\begin{center}
$A^{t\bar{t}}_\mathrm{FB} =$ (16 $\pm $ 4.5)\%, while the  SM NNLO prediction is $A^{t\bar{t}}_\mathrm{FB} =$ (9.5 $\pm $ 0.7)\% \cite{czakon_tev-afb}.
\end{center}

\begin{wrapfigure}{r}{0.40\textwidth}
\vspace{-5mm}
\centering
\begin{tabular}{c}
\includegraphics[width=0.39\textwidth]{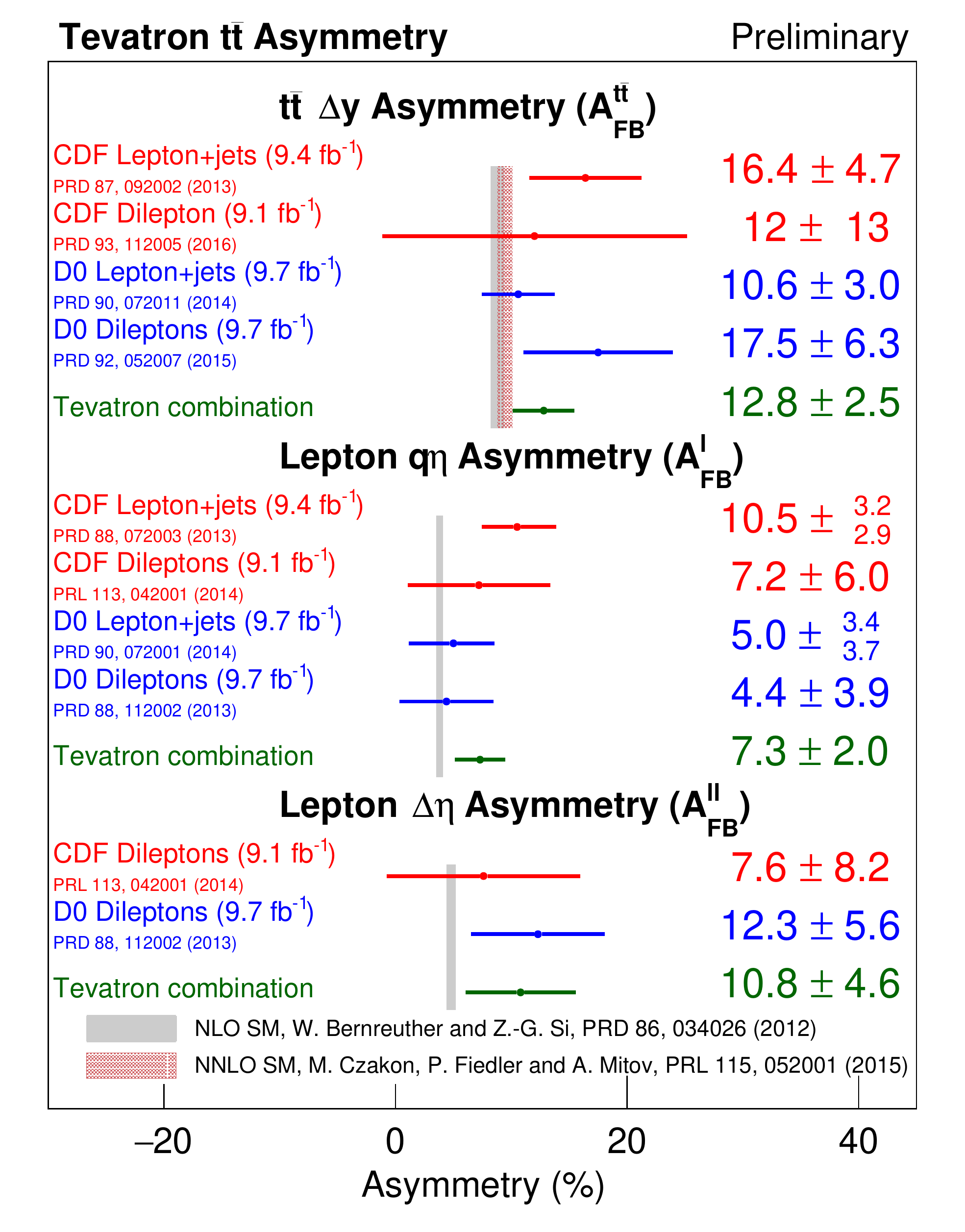}
\end{tabular}
\vspace{-6mm}
\caption{{\small Summary of the Tevatron FB asymmetry measurements.}}
\label{fig:Afb_sum_Tev}
\vspace{-3mm}
\end{wrapfigure}
\noindent In addition, the measured $A^{t\bar{t}}_\mathrm{FB}$ as a function of the top-quark pair invariant mass $M_{t\bar{t}}$ gives systematically higher values in comparison with the SM NLO prediction. A tension between the SM prediction and the measurements is evident but not big.

The D0 experiment measured the FB asymmetry in the $\ell+$jets channel using the data sample of 9.7~fb$^{-1}$ \cite{prd90_d0_afb}. The $t\bar{t}$ rapidity difference, $\Delta{y}$,  was employed to extract the asymmetry. 
The measured asymmetry unfolded to parton level is $A^{t\bar{t}}_\mathrm{FB} =$ (10.6 $\pm $ 3.0)\%. 
The FB asymmetry was investigated also as a function of $\Delta y$ and  $M_{t\bar{t}}$. 
The measured value of $A^{t\bar{t}}_\mathrm{FB}$ is in agreement with the SM prediction, though the dependence of $A^{t\bar{t}}_\mathrm{FB}$ on $\Delta y$ is systematically higher than the expected theoretical dependence.
In the $\ell\ell$ channel, the D0 experiment measured the $A^{t\bar{t}}_\mathrm{FB}$ asymmetry along with the top-quark polarization $\kappa P$ in the beam axis using the data sample of 9.7~fb$^{-1}$ \cite{prd92_d0_afb-ll}. 
The $A^{t\bar{t}}_\mathrm{FB}$ was measured using $\Delta y$ and polarization $\kappa P$ was extracted from the distribution of $\ell^{+}$ and  $\ell^{-}$:
 $d\sigma/d\mathrm{cos} \theta^{\pm} = (1+\kappa P \mathrm{cos} \theta ^{\pm })/2$.
The obtained results are 
\begin{center}
\vspace{-2mm}
$A^{t\bar{t}}_\mathrm{FB}$ = 15.0 $\pm $ 6.4 (stat) $\pm $ 4.9 (syst)\% and $\kappa P =$ 7.2 $\pm $ 10.5 (stat) $\pm $ 4.2 (syst)\%.
\vspace{-2mm}
\end{center}
\noindent The result for the $\ell\ell$ channel was combined 
(see Ref. \cite{prd92_d0_afb-ll}) with that of the $\ell+$jets channel resulting in
$A^{t\bar{t}}_\mathrm{FB}$ = (11.8 $\pm $ 2.8) \%,
which agrees  with the SM prediction.

\noindent {\bf Top-quark pair leptonic asymmetries.} 
The leptonic top-quark asymmetries can be 
defined in the $\ell+$jets channel ($A^{\ell}_\mathrm{FB}$) as well as in the $\ell\ell$ channel ($A^{\ell}_\mathrm{FB}$ and $A^{\ell\ell}_\mathrm{FB}$):
\vspace*{-8pt}
\begin{eqnarray} 
A^{\ell}_\mathrm{FB}     = \frac{N_\mathrm{qy_{\ell}>0}-N_\mathrm{qy_{\ell}<0}}{N_\mathrm{qy_{\ell}>0}+N_\mathrm{qy_{\ell}<0}}, 
A^{\ell\ell}_\mathrm{FB} = \frac{N_\mathrm{\Delta \eta >0}-N_\mathrm{\Delta \eta <0}}{N_\mathrm{\Delta \eta >0}+N_\mathrm{\Delta \eta <0}}, 
\label{eq:AFB_lep}
\end{eqnarray}
\vspace*{-12pt}

\noindent where $q\ (y_{\ell})$ is the lepton charge (rapidity), $\Delta \eta =\eta_{\ell^{+}}-\eta_{\ell^{+}}$ is the pseudorapidity difference of $\ell\ell$ pair,  and  $N_{x<0(>0)}$ is the corresponding number of events.
The individual CDF and D0  measurements of the leptonic asymmetries are summarized in Fig.\ref{fig:Afb_sum_Tev}.
\noindent {\bf Combination of the $t\bar{t}$ asymmetries.} 
The CDF and D0 experiments combined their measurements of the $A_\mathrm{FB}^{t\bar{t}}$ (reconstructed using $\Delta y$) and leptonic asymmetries ($A_\mathrm{FB}^{l}$, 
$A_\mathrm{FB}^{ll}$)  and compared the results with the SM NLO and NNLO predictions \cite{cdf-d0_comb_l6}. The combination of 
the asymmetry $A_\mathrm{FB}^{t\bar{t}}$ gave:
$A^{t\bar{t}}_\mathrm{FB}$ = (12.8 $\pm $ 2.5) \%, 
which is within 1.5 $\sigma $ of the SM NNLO prediction.

\vspace*{-12pt}
\subsection{Polarization of $W$ bosons from top-quark decays}

Under the SM the top-quark decays before hadronization through the weak interaction mainly to $Wb$. It enables tests of the V-A structure of the electroweak interactions by studying the  $tWb$ coupling and $W$-boson polarization. The SM prediction for the longitudinal, 
left- and right-handed polarizations is $f_{0}$ = 0.696, $f_{-}$ = 0.303   and   $f_{+}$  = 3.8$\times $10$^{-4}$, respectively. 

CDF performed a model-independent measurement in the $\ell+$jets channel using a dataset of 9.1~fb$^{-1}$ \cite{cdf_prd87}. The result of simultaneously determined $f_{0}$ and $f_{+}$ is
\begin{center}
\vspace{-2mm}
$f_{0}$ = 0.726 $\pm $ 0.066(stat) $\pm $ 0.067(syst), $f_{+}$ = -0.025 $\pm $ 0.044(stat) $\pm $ 0.058(syst). 
\vspace{-2mm}
\end{center}
Fixing $f_{+}$ and $f_{0}$ at their SM values, the CDF obtained:
\begin{center}
\vspace{-2mm}
$f_{0}$ = 0.683 $\pm $ 0.042(stat) $\pm $ 0.040(syst), $f_{+}$ = -0.025 $\pm $ 0.024(stat) $\pm $ 0.040(syst). 
\vspace{-2mm}
\end{center}
So the measured polarization is in good agreement with the SM prediction.

\vspace*{-10pt}
\subsection{Top-quark polarization and spin correlations}
The SM predicts that top quarks produced at the Tevatron collider are almost unpolarized,
while some of BSM models predict enhanced polarizations \cite{fayfer2012}. D0 measured the top-quark polarization along three quantization axes (beam, helicity and transverse axes) in the $\ell+$jets channel using the full data set of 9.7~fb$^{-1}$ \cite{d0_top-pol}. The polarizaton was measured using distributions of leptons along the aforementioned axes:
\vspace*{-8pt}
\begin{equation}
\frac{1}{\Gamma }\frac{d\Gamma }{dc_{\theta_{1}}dc_{\theta_{2}}} = \frac{1}{4}\left(1+\kappa_{1}P_{\vec{n}}c_{\theta_{1}} 
            +\rho \kappa_{2}P_{\vec{n}}c_{\theta_{2}} - \kappa_{1}\kappa_{2}Cc_{\theta_{1}}c_{\theta_{2}}   \right),
\label{eq:top_pol}
\end{equation}
\vspace*{-12pt}

\noindent where $c_{\theta_{i}}$ is the cosine of the $i^\mathrm{th}$ lepton production angle  $\theta_{i}$ with respect to axis $\vec{n} $, $P_{\vec{n}}$ is the polarization with respect to axis $\vec{n} $, $C$ is the spin correlation coefficient and $\kappa_{i}$ is the spin analyzing power \cite{uwer2002}. 
The measured polarizations for 
 the beam ($P_\mathrm{bea}$), helicity ($P_\mathrm{hel}$) and transverse ($P_\mathrm{tra}$) axes are:
 
  $P_\mathrm{bea}$ = +0.081 $\pm $ 0.048 , $P_\mathrm{hel}$ = 0.102 $\pm $ 0.061 , and $P_\mathrm{tra}$ = +0.040 $\pm $ 0.034. Correspondingly the SM predictions are: -0.002, -0.004 and +0.011.

{\bf Top-quark spin correlations.} The top-quark lifetime, $\tau_\mathrm{top}\approx 5\cdot 10^{-25}\,s$ is much shorter than the spin-decorrelation time, $\tau_\mathrm{spin}\approx 3\cdot 10^{-21}\,s$. It means that the production spin characteristics are transferred to the top-quark decay products without dilution. 
The SM predicts that $t$  and $\bar{t}$ are produced practically unpolarized but their spins are correlated.  
A spin correlation observable can be defined as
\vspace*{-8pt}
\begin{equation}
O     = \frac{\sigma(\uparrow \uparrow )+\sigma(\downarrow \downarrow)-\sigma(\uparrow \downarrow)-\sigma(\downarrow \uparrow)}
             {\sigma(\uparrow \uparrow )+\sigma(\downarrow \downarrow)+\sigma(\uparrow \downarrow)+\sigma(\downarrow \uparrow)}
\label{eq:spinCorr_d0}
\end{equation}
\vspace*{-15pt}

\noindent where  $\sigma $ is the $t\bar{t}$ production cross section and the arrows refer to the spin states of the $t$ and $\bar{t}$ quarks relative to their quantization axes.

D0 measured the $t\bar{t}$ spin correlation strength 
using the data set corresponding to an integrated luminosity of 9.7 fb$^{-1}$ \cite{spinCorr_d0}. 
The measurement was performed in the $\ell+$jets and $\ell\ell$ channels. In the analysis the off-diagonal spin basis maximizing correlation for $p\bar{p}$ was employed. The obtained results read:
\begin{center}
\vspace{-1mm}
$O_\mathrm{off}$ = 0.89 $\pm $ 0.16(stat) $\pm $ 0.15(syst), while the SM value is $O_\mathrm{off}$ = 0.80 $^{+0.01}_{-0.02}$. 
\end{center}
The significance of the measured strength from zero is 4.2~$\sigma $. Assuming absence of non-SM physics, the fraction of $gg$ fusion at the $t\bar{t}$ production was extracted: $f_{gg}$ = 0.08 $\pm $ 0.16 which is in good agreement with the SM expectation $f^\mathrm{SM}_{gg}$ = 0.135.

\vspace*{-10pt}
\subsection{Other top-quark properties}
{\bf Top-quark decay width.} In the SM the top-quark decay width has been calculated at NNLO in QCD and assuming the top-quark mass, $m_\mathrm{top}$ =172.5 GeV/$c^{2}$, its value is $\Gamma _\mathrm{top}$ = 1.32 GeV \cite{Gao2013}.

D0 determined $\Gamma _\mathrm{top}$ using a model-dependent indirect measurement  that assumes SM couplings. The width was determined in a data set corresponding to an integrated luminosity of 5.4~fb$^{-1}$ and the obtained value was $\Gamma _\mathrm{top}$ = 2.00$^{+0.47}_{-0.43}$ GeV \cite{top-width_d0}.
 
 CDF employed a more model-independent measurement  based on  a direct shape comparison of the reconstructed $m_\mathrm{top}$ distribution in data to  the simulated top-quark mass distributions. The analysis was carried out in the $\ell+$jets channel with a data set of 8.7~fb$^{-1}$ \cite{top-width_cdf}. 
A likelihood fit was applied to data to  extract $\Gamma _\mathrm{top}$: \begin{center}
\vspace{-2mm}
$\Gamma_\mathrm{top}$ = 1.63 GeV 
 and 1.10  $< \Gamma_\mathrm{top} <$ 4.05 GeV 68\% C.L.. 
\end{center}

{\bf Top-quark charge.}
Determination of the top-quark charge was initiated mainly  by a need  to confirm that the top quark is really the top quark of the SM decaying into $W^+$ boson and $b$ quark ($t\rightarrow W^{+} + b$) and not an exotic quark with the charge of -4/3 decaying under the scheme: $t\rightarrow W^{-} + b$.

The CDF and D0 experiments investigated the top-quark charge in data sets of 5.6~fb$^{-1}$ \cite{top-charge_cdf} and 5.3~fb$^{-1}$ \cite{top-charge_d0}, respectively. 
The observables used for the top-quark charge determination are constructed from the charge of the lepton ($Q_{\ell}$) and $b$-jet charge in the leptonic ($Q_{b}^{\ell}$) and hadronic ($Q^\mathrm{h}_{b}$) branches. 
The CDF experiment used $Q_\mathrm{comb}=Q_{\ell}\times Q_{b}$ and by performing a statistical analysis 
based on the pseudoexperiments 
excluded the exotic quark hypothesis at the 99\% confidence level. The D0 experiment defined 
$Q^{\ell}_{t}= |Q_{\ell}+Q^{\ell}_{b}|$ and $Q^\mathrm{h}_{t}= |-Q_{\ell}+Q^\mathrm{h}_{b}| $ and excluded the hypothesis claiming that all top quarks in the data are  exotic quarks, at more than 5$\sigma $.

\vspace*{-12pt}
\section{Conclusion}
The Tevatron experiments, CDF and D0, provided us with remarkable results on top-quark properties which are in good agreement with the SM.
At the Tevatron,  not only the basic characteristics of the top quark like the cross section or top-quark mass have been measured,  but also such quantities as
the asymmetries in $t\bar{t}$ production, the top-quark spin correlations, the helicity of $W$ bosons coming from top-quark decays, the top-quark charge, its decay width, etc.
Some of the results like the production asymmetries or the spin correlations are unique due to fact that $p\bar{p}$ collisions are to some extent complementary to $pp$ collisions at the LHC.
The results from CDF and D0 are still coming and  keep a high level of quality.


\vspace*{-12pt}
 
\end{document}